\documentclass[reprint,amsmath,amssymb]{revtex4-1}

\usepackage{subfigure}
\usepackage{dcolumn}
\usepackage{bm}
\usepackage[pdftex]{graphicx}

\begin{document}
\title{Direct evaluation of free energy for large system  \\ through structure integration approach}
\author{Kazuhito Takeuchi}
\author{Ryohei Tanaka}
\author{Koretaka Yuge}
\affiliation{Department of Materials Science and Engineering, Kyoto University, Sakyo, Kyoto 606-8501, Japan}

\date{\today}
\pacs{1}

\begin{abstract}
We propose a new approach, "structure integration", enabling direct evaluation of configurational free energy for large systems. The present approach is based on the statistical information of lattice. Through first-principles-based simulation, we find that the present method evaluates configurational free energy accurately in disorder states above critical temperature.
\end{abstract}

\maketitle

\section{\label{sec:level1} Introduction}
In the present study, we are interested in expressing microscopic states for substitutional crystalline solids, which is described in terms of parameters independent of system (e.g., constituent elements). In order to apply this expression to calculating macroscopic property, we choose a topic to calculate configurational free energy in alloy system as a model case. Helmholtz free energy, $F$, is derived from partition function, $Z$, which is defined by:
\begin{equation}
\label{eq:exact}
Z= \sum_{E} W(E) \exp \left(-\frac{E}{k_{\rm B}T} \right) \simeq \sum_{s} \exp \left(-\frac{E(s)}{k_{\rm B}T} \right),
\end{equation}
where $W$ is the number of states, $s$ denotes atomic arrangement, $k_{\rm B}$ is boltzmann constant and $E(s)$ is total energy of $s$. The last equation is allowed when microscopic states are confined to atomic arrangements on a given lattice, and $\sum_s$ represents sum over all possible atomic arrangements\cite{solid-state}. For large system, exact calculation of $Z$ is practically difficult because we do not know $W$ {\it a priori}. In order to overcome this difficulty, several successful approaches and techniques have been proposed and widely used such as cluster variational method (CVM)\cite{PhysRev.81.988, :/content/aip/journal/jcp/60/3/10.1063/1.1681115} and efficient Monte Carlo (MC) simulations\cite{:/content/aip/journal/jcp/21/6/10.1063/1.1699114,PhysRevLett.61.2635,PhysRevLett.68.9,PhysRevLett.71.211}.

When temperature, $T$, increases, the probability of high energy state increases and entropy comes to contribute to the macroscopic properties. Then the system can go into disorder states. In disorder states, direct estimation of property is nontrivial because possible states mainly contributing to property increase with increase of system size. In our research group, alloy disorder state is successfully described by statistical information of lattice\cite{Yuge_sro}. Based on this theory, we propose a new approach, "structure integration", to evaluate configurational free energy directly in large binary alloy system. We successfully give analytical representation of $W$ in terms of the so-called "correlation functions"\cite{SDG} which does not depend on constituent elements. Finally we apply present method to Cu-Au binary alloy system and confirm its validity and applicability.

\section{\label{sec:level2} methodogy}
We introduce Ising-like spin variable $\theta_i$ that specifies the occupation of element on lattice site $i$ (e.g., in A-B binary system, $ \theta_i = +1 $ for A and $ \theta_i = -1 $ for B at site $i$). Using cluster expansion\cite{SDG} (CE), we can obtain correlation functions, $\xi_k$, that completely represent atomic arrangement where $k$ specifies the cluster that consists of lattice points (e.g., nearest neighbor pair, triangle, and second nearest neighbor pair). We can expand configurational property (energy in the present study) using $\xi_k$, namely,
\begin{equation}
\label{eq:CE}
E = \sum_{k} V_{k} \xi_{k} ,
\end{equation}
where $V_k$ is called effective cluster interaction (ECI). $V_k$ is the coefficient of $\xi_k$ in order to represent $E$, i.e., $V_k$ is the projection of $E$ onto $\xi_k$ and can be practically determined from first principle calculation.  In this expression, we separate the variables that depend on constituent elements, $V_k$s, and are independent of those, $\xi_k$s. Applying Eq.~(\ref{eq:CE}) to Eq.~(\ref{eq:exact}), we can get the expression of $Z$ that are explicitly separated into the number of states in terms of $\xi_k$s that is independent of constituent elements and the exponential function including the parameters that are depend on constituent elements. In following discussion, in order to perform Eq.~(\ref{eq:exact}) practically for large system, we read $W$ as the atomic density of states (DOS) that is defined as the function in terms of total energy for atomic arrangement, which naturally leads to replacing discrete summation, $\sum$, by integration, $\int$.

\begin{figure}
  \includegraphics[clip, width=\columnwidth]{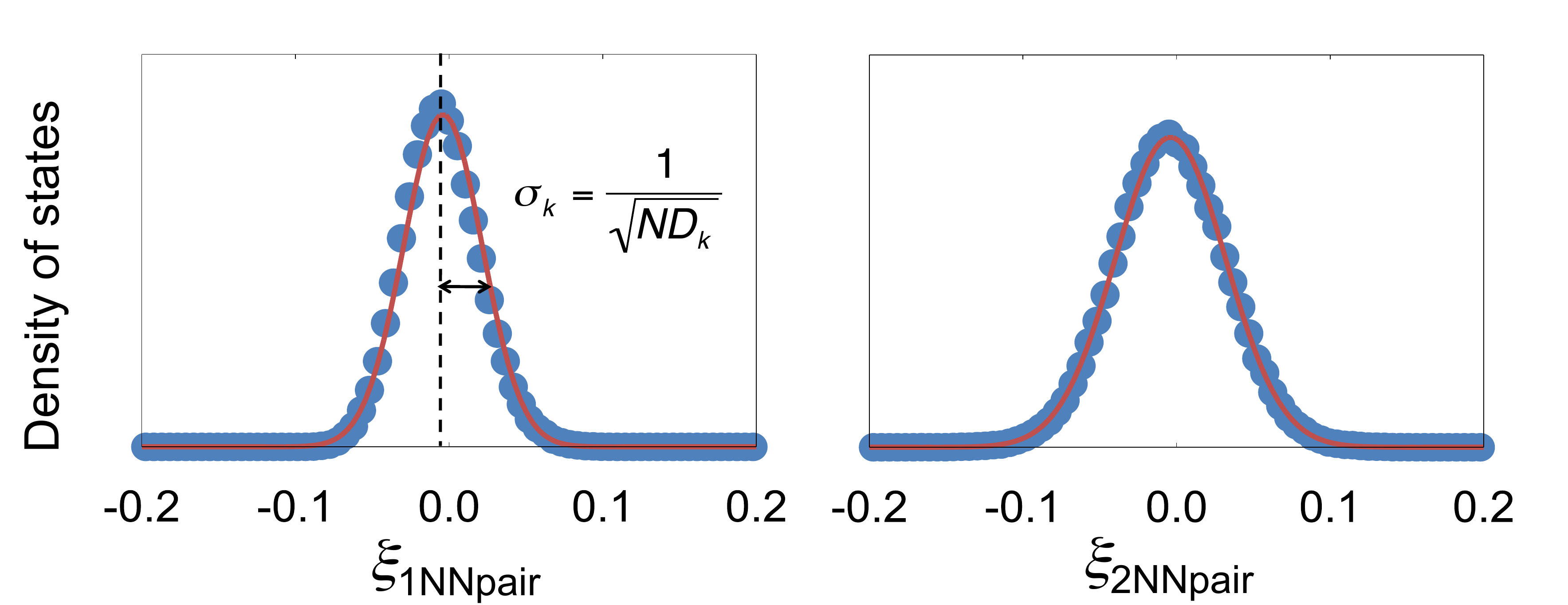}
  \caption{ Density of states in terms of correlation function. Points and solid curves denote the results of MC simulation in $\rm{A_{128}B_{128}}$ binary system on fcc lattice and normal distribution.}%
  \label{fig:corr_dist}%
\end{figure}

In A-B binary system, $\xi_k$ is equivalent to the sum of Ising-like spin product over all the $k$ type clusters on underlying lattice, i.e., $\xi_k = \sum_{k \in s} ( \prod_{i \in k} \theta_i ) / \sum_{k \in s} 1$ where $i$ denotes the lattice point in $k$ and  $\sum_{k \in s}$ denotes that summation taken over symmetry-equivalent clusters to figure $k$ in given atomic arrangement, $s$. Since $\xi_k$ is a contribution from an atomic arrangement to $k$, the DOS in terms of $\xi_k$ also represents the marginal distribution of the DOS in terms of all possible $\xi_k$s. It has been shown that when number of atoms in the system increases the DOS in terms of $\xi_k$ is given by the normal distribution function\cite{Yuge_sro} whose average is $ \mu_k = (2x-1)^k$ and standard deviation is $ \sigma_k = 1/\sqrt{ND_k}$ where $D_k$ is the number of cluster $k$ per site\cite{PhysRevB.42.9622}. For instance, in $ \rm{A}_{128} \rm{B}_{128}$ binary system on fcc lattice, we show in Fig.~\ref{fig:corr_dist} the analytical normal distributions and the histograms of $\xi_k$ where $k$ is first nearest neighbor pairs (1NN) and 2NN pairs derived from MC simulation. Atomic arrangements with $\xi_k \simeq 0$ are majority, corresponding to ideally disorder states, and those with $\xi_k$ far from $0$ are minority, typically corresponding to well-ordered state such as $L1_0$. In the above we focus on the single variable DOS, but the atomic DOS, $W$, obeys to the multivariate distribution function in terms of $\xi_k$s. In order to express the form of $W$, we should know not only all the marginal distribution of $W$ but also correlation coefficients between different $\xi_k$s. The off-diagonal elements of correlation coefficient matrix, \mbox{\boldmath $R$}, correspond to the correlation coefficients between different $\xi_k$s. We confirm that the off-diagonal elements of \mbox{\boldmath $R$} are approximately zero when size increases (Fig.~\ref{fig:R}) using MC simulation (when $x \neq 0.5$ or in multicomponent system, this is not always true. See Appendix~\ref{sec:appendixB}.). This indicates that when we consider large system, off-diagonal elements of \mbox{\boldmath $R$} can be neglected. Therefore when we describe $W$ in terms of all of the $\xi_k$s as the multivariate normal distribution function, $P ( \xi_{\alpha} , \xi_{\beta},... )$, it can be decomposed into the product of the normal distribution function, $P(\xi_k)$\cite{Yuge_sro},
\begin{equation}
\label{eq:decompose}
P ( \xi_{\alpha} , \xi_{\beta},... ) \simeq \prod_{k} P_{k}(\xi_{k}).
\end{equation}
Therefore we can e the partition function $Z$ as
\begin{equation}
\label{eq:SI}
Z \simeq A \prod_{k} \int P_{k}(\xi_{k}) \exp \left( -\frac{V_{k} \xi_{k}}{k_{\rm B}T} \right)d\xi_k .
\end{equation}
Here $A$ is normalization constant for integration. This integration is easy to perform when we know a set of $V_k$s, and we call this equation "structure integration" since integrating variable, $\xi_k$, denotes structure. When the number of $k$ is $1$ in Eq.~(\ref{eq:CE}), $E=V\xi$, as a simple case, it is easily shown that $Z$ in terms of $E$ is equivalent to $Z$ in terms of $\xi$ (see Appendix~\ref{sec:appendix}). Note that the present approach is essentially different from high-temperature series expansions. In high-temperature series expansion, $Z$ is expanded under the condition $J/k_{\rm B}T \to 0$ , where $J$ is interaction. However, in the present method, we neglect the correlation coefficients between different correlation functions in Eq.~(\ref{eq:decompose}) but do not suppose $J/k_{\rm B}T \to 0$. The accuracy and validity of present method rely only on lattice. This simple modeling does not need computational costs which is needed in MC simulation. Since we successfully take the variables that are independent of system apart from $Z$, we can evaluate configurational free energy directly for specific $V_k$s.
Note that in particular at compositions far from $x=0.5$, the DOS in terms of $\xi_k$ cannot be well expressed by the normal distribution function.
This problem is briefly discussed in Appendix~\ref{sec:appendixB}.

\begin{figure}
  \includegraphics[clip, width=\columnwidth]{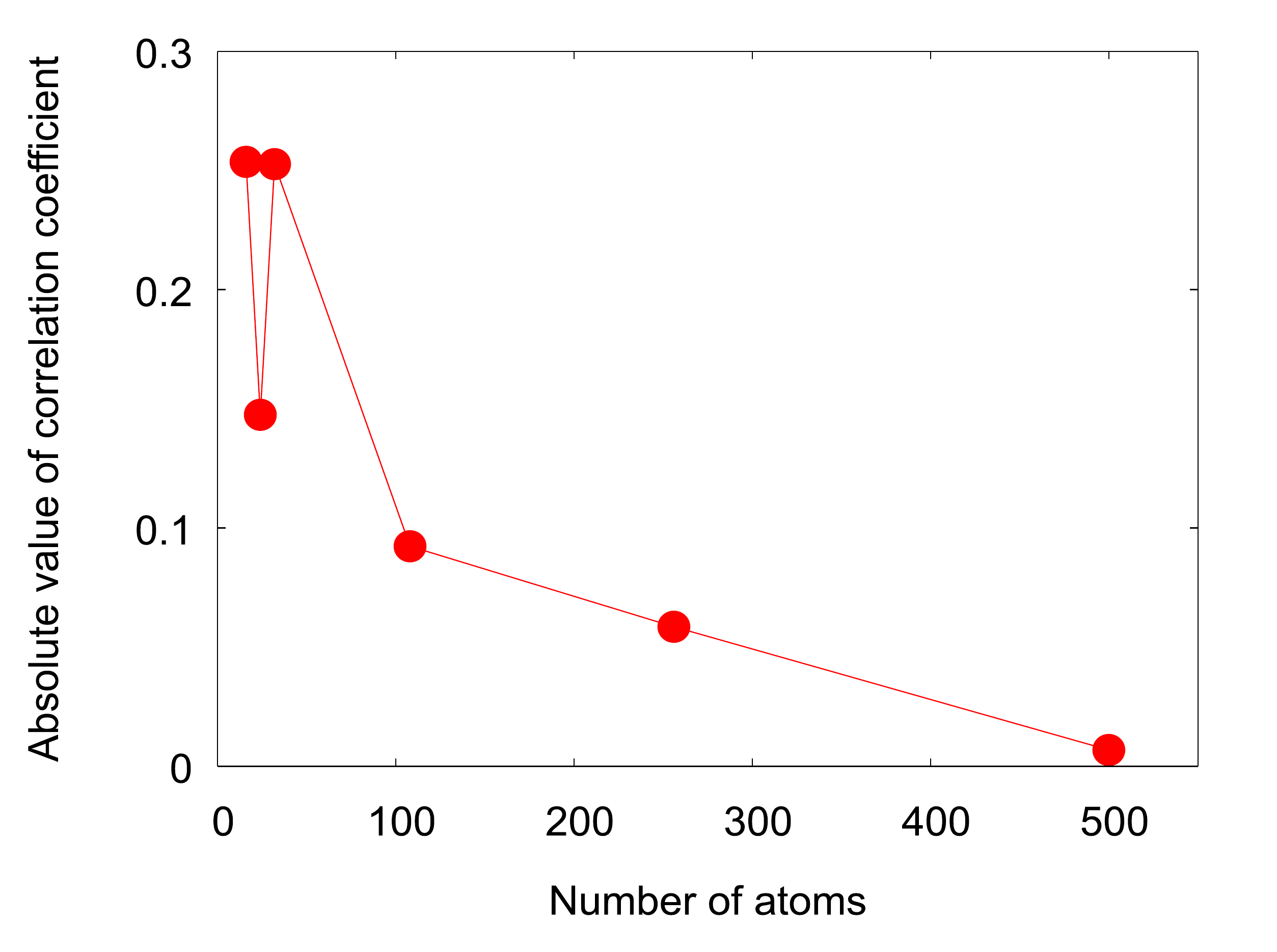}
  \caption{Absolute of correlation coefficient between $\xi_{\rm 1NNpair}$ and $\xi_{\rm 2NNpair}$ as a function of number of atoms in the system.
  }%
  \label{fig:R}%
\end{figure}

Here we emphasize the difference between our method and CVM because both methods are based on $\xi_k$s. Therefore, the concept of our method seems to be similar to that of CVM.
However our method is essentially different from CVM.
We can see this essential difference through the process of estimating free energy. In CVM, $F$ in equilibrium states is determined by minimizing $F$ with respect to $\xi_k$s at specific $T$ through numerical method such as natural iteration method\cite{:/content/aip/journal/jcp/60/3/10.1063/1.1681115}. This is also interpreted as in CVM, $F$ in equilibrium states being expressed by ensemble averaged $\xi_k$s at specific $T$ through the optimization of $\xi_k$s. In contrast, our method allows us direct estimation of $F$ at any $T$. This is because our method rewrites the DOS over all configurational space in terms of $\xi_k$s. Namely, $F$ in CVM is described by the ensemble averaged $\xi_k$s at specific $T$, while $F$ in our method is described by the DOS in terms of $\xi_k$s over all configurational space through the approximation in Eq.~(\ref{eq:decompose}).

In order to calculate Eq.~(\ref{eq:SI}), we should determine $V_k$s that are only the variables depending on constituent elements. $V_k$s are obtained by applying DFT total energies for multiple atomic arrangements, $E(s)$ to Eq.~(\ref{eq:CE}). Total energies are obtained by the first-principles calculation using the VASP code\cite{PhysRevB.47.558,PhysRevB.54.11169}, based on the projector-augmented wave method (PAW)\cite{PhysRevB.59.1758} within the generalized-gradient approximation of Perdew-Burke-Ernzerhof (GGA-PBE)\cite{PhysRevLett.77.3865} to the exchange-correlation functional. The plane wave cutoff of 400 eV is used, and atomic positions are kept fixed on underlying fcc lattice. In the present study, we chose CuAu binary alloy as model system.  Total energies of 223 structures consist of up 32 atoms are calculated by DFT.

\begin{figure}
  \includegraphics[clip, width=\columnwidth]{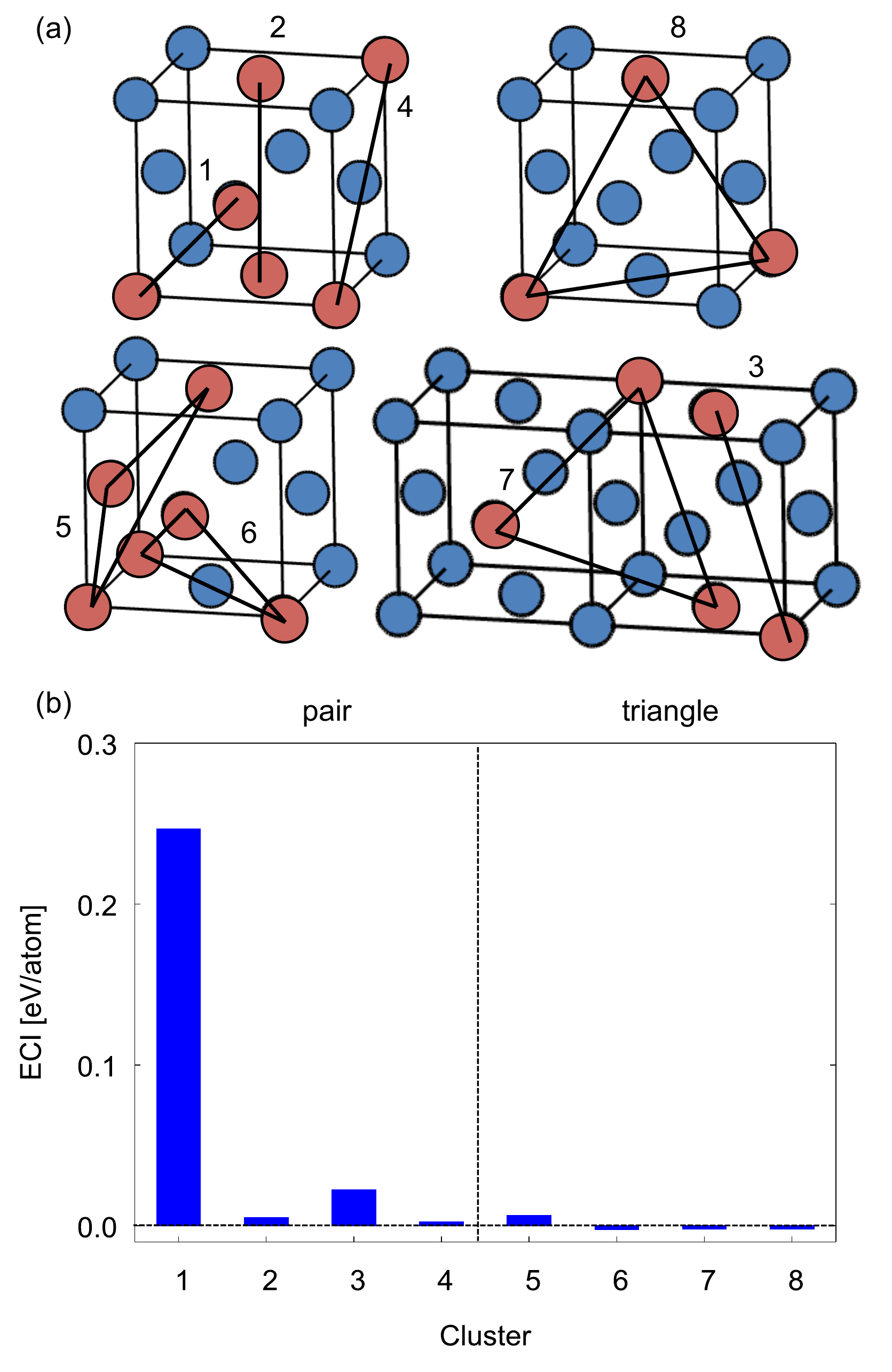}
  \caption{
  (a) Schematic illustration of multibody clusters used in the CE. (b) ECI for the $8$ multibody clusters.}%
  \label{fig:eci}%
\end{figure}

\section{\label{sec:level3} Results and discussion}
We obtained nine optimized ECI with prediction accuracy, a cross-validation score\cite{vandeWalle2002539}, of 0.75 meV/atom, which gives sufficient accuracy to capture the thermodynamics for CuAu alloys (Fig.~\ref{fig:eci}). Dominant contribution comes from cluster 1, i.e., nearest neighbor pair. The pair ECI exhibits positive sign, indicating strong preference of unlike-atom pair along this coordination.

\begin{figure}
  \includegraphics[clip, width=\columnwidth]{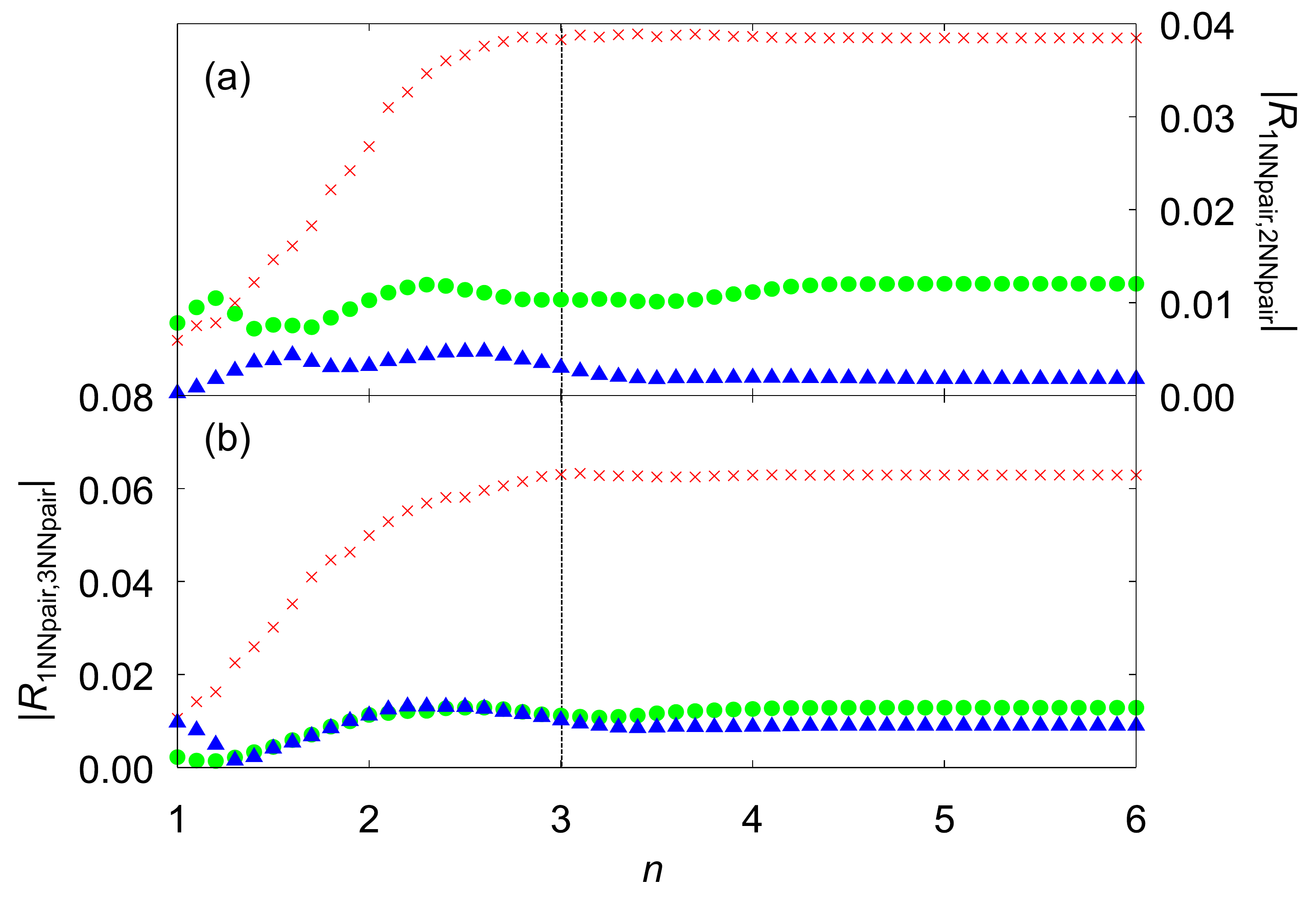}
  \caption{The absolute value of correlation coefficient between (a) $\xi_{\rm 1NNpair}$ and $\xi_{\rm 2NNpair}$ and (b) $\xi_{\rm 1NNpair}$ and $\xi_{\rm 3NNpair}$ under the condition $ |\xi_k| < n \sigma_k$ by MC simulation. Cross marks, circles and triangles denote correlation coefficients when $N=256, 4000, 6912$. Broken line denotes $\sigma_k = 3$. }
  \label{fig:coefvsvar}%
\end{figure}

First of all in order to calculate Eq.~(\ref{eq:SI}), the intervals of integration should be determined. Since the variables of integration correspond to atomic arrangement, the intervals of integration should not include unreal atomic arrangements, thus should be limited to the range of good  agreement between $W$ and Eq.~(\ref{eq:decompose}). For example in $\xi_{\rm 1NNpair}$ on fcc lattice, the minimum value of $\xi_{\rm 1NNpair}$ is $-1/3$ that corresponds to an order state, $L1_0$. However the minimum value of the interval of integration, $\xi_{\rm 1NNpair}$, should not be $-1/3$. For example in $L1_0$, when $\xi_{\rm  1NNpair}$ is changed, $\xi_{\rm 2NNpair}$ is also explicitly changed. Eq.~(\ref{eq:decompose}) requires that all correlation functions are nearly independent, however in $L1_0$ $\xi_{\rm  1NNpair}$ and $\xi_{\rm 2NNpair}$ are nearly linear dependent. Likewise in order and partially order states, some of $\xi_k$s are not considered as independent. Thus order and partially order states do not satisfy the requirement for Eq.~(\ref{eq:decompose}), and the normal distribution function cannot appropriately describe order and partially order states. This is the reason why including order and partially order states in the intervals of integration in Eq.~(\ref{eq:SI}) leads to non-physical result and determining the intervals of integration is important. In Fig.~\ref{fig:coefvsvar}, we sample sufficient atomic arrangements with MC simulation and plot the correlation coefficients versus sampling range, $n \sigma_k$. We do not use $\xi_k$s but $\sigma_k$s in order to determine the interval of integration because $\sigma_k$s are only quantitative values when we describe the DOS in terms of $\xi_k$ as the normal distribution function. Correlation coefficient becomes large as $n$ increases, and when $n \le 3$ all the absolute values of correlation coefficients appears to converge. This means that almost all the atomic arrangements is considered when $n<3$. Thus $n=3$ is considered as the boundary value required for Eq.~(\ref{eq:decompose}). Atomic arrangements in $n \le 3$ satisfy the requirement for Eq.~(\ref{eq:decompose}) so we should consider the intervals of integration as $-3\sigma_k \le \xi_k \le 3\sigma_k$.

\begin{figure}
  \includegraphics[clip, width=\columnwidth]{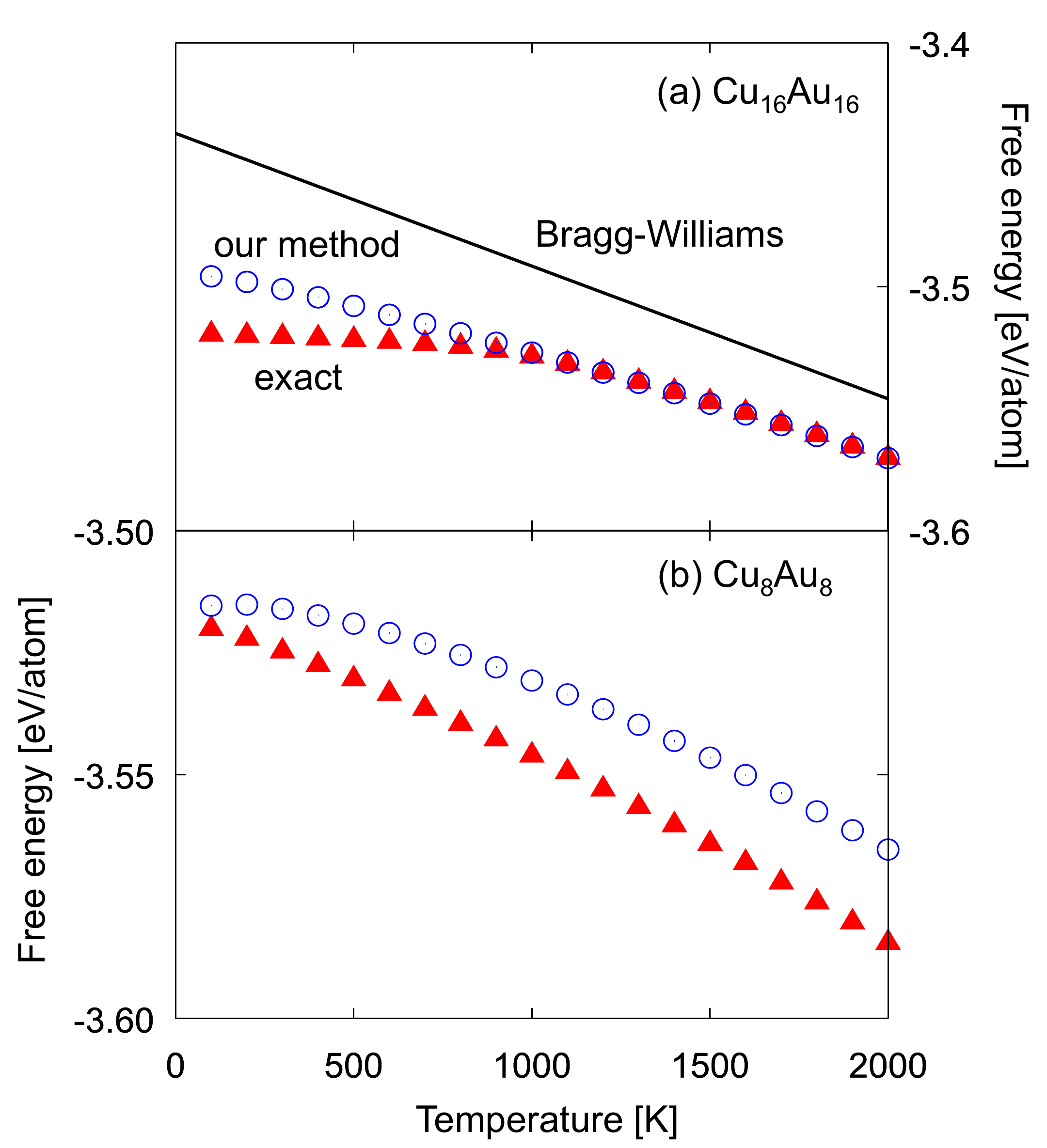}
\caption{Calculated free energy for equiatomic Cu-Au alloys. Closed triangles and open circles denote exact result from Eq.~(\ref{eq:exact}) and the result through present method. In Fig.~\ref{fig:result}(a), solid line denotes free energy using Bragg-Williams approximation. (a) Free energy for $\rm{Cu}_{16}\rm{Au}_{16}$ and (b) for $\rm{Cu}_{8}\rm{Au}_{8}$.}
\label{fig:result}%
\end{figure}

In $\rm{Cu}_{16}\rm{Au}_{16}$ system with the $2 \times 2 \times 2$ expansion of fcc unit cell, we calculate total energy for all possible states and obtain exact free energy from Eq.~(\ref{eq:exact}). The number of all possible states is $_{32}\rm{C}_{16} \simeq 6 \times 10^{8}$ so this size is the limit to calculate exactly because of computational costs. Integrating Eq.~(\ref{eq:SI}), we obtain free energy and compare with the exact result in Fig.~\ref{fig:result}. This result shows that present method describes successfully the exact configurational free energy when $T$ is larger than order-disorder transformation temperature, $T_{\rm c} = 950$[K]. $T_{\rm c}$ is determined by MC simulation and ECI.
In order to compare the predction accuracy of our method with that of Bragg-Williams approximation, we also show the free energy using Bragg-Williams approximation, $F = V_0 - k_{\rm B}T \ln W $, described by solid line in Fig.~\ref{fig:result}(a),
where $V_0$ is empty cluster ($V_0$ means the enthalpy in perfectly disorder states) and $W$ is the number of states, $_{32}\rm{C}_{16}$ (when $N \to \infty$, $W$ becomes $N(x \ln x + (1-x) \ln (1-x))$ because of Stirling’s approximation).
Since $V_0$ becomes negligible with respect to $F$ when $T \to \infty$, free energies in Fig.~\ref{fig:result}(a) become numerically equall to $-k_{\rm B}T \ln W$.
Although Bragg-Williams approximation successfully reproduce $T$-dependence of $F$ at sufficiently high $T$, we can see significant deviation in absoulte value of $F$ compared with the exact result.
Meanwhile Fig.~\ref{fig:result}(a) clearly shows that between 950-2000[K] our method can predict free energy better than Bragg-Williams approximation. This is because Bragg-Williams approximation cannot essentially include the contribution from short-range order to free energy, while our method can include such contribution through multibody interaction such as $V_{\rm 1NNpair}$ and $V_{\rm 2NNpair}$.

\begin{figure}
  \includegraphics[clip, width=\columnwidth]{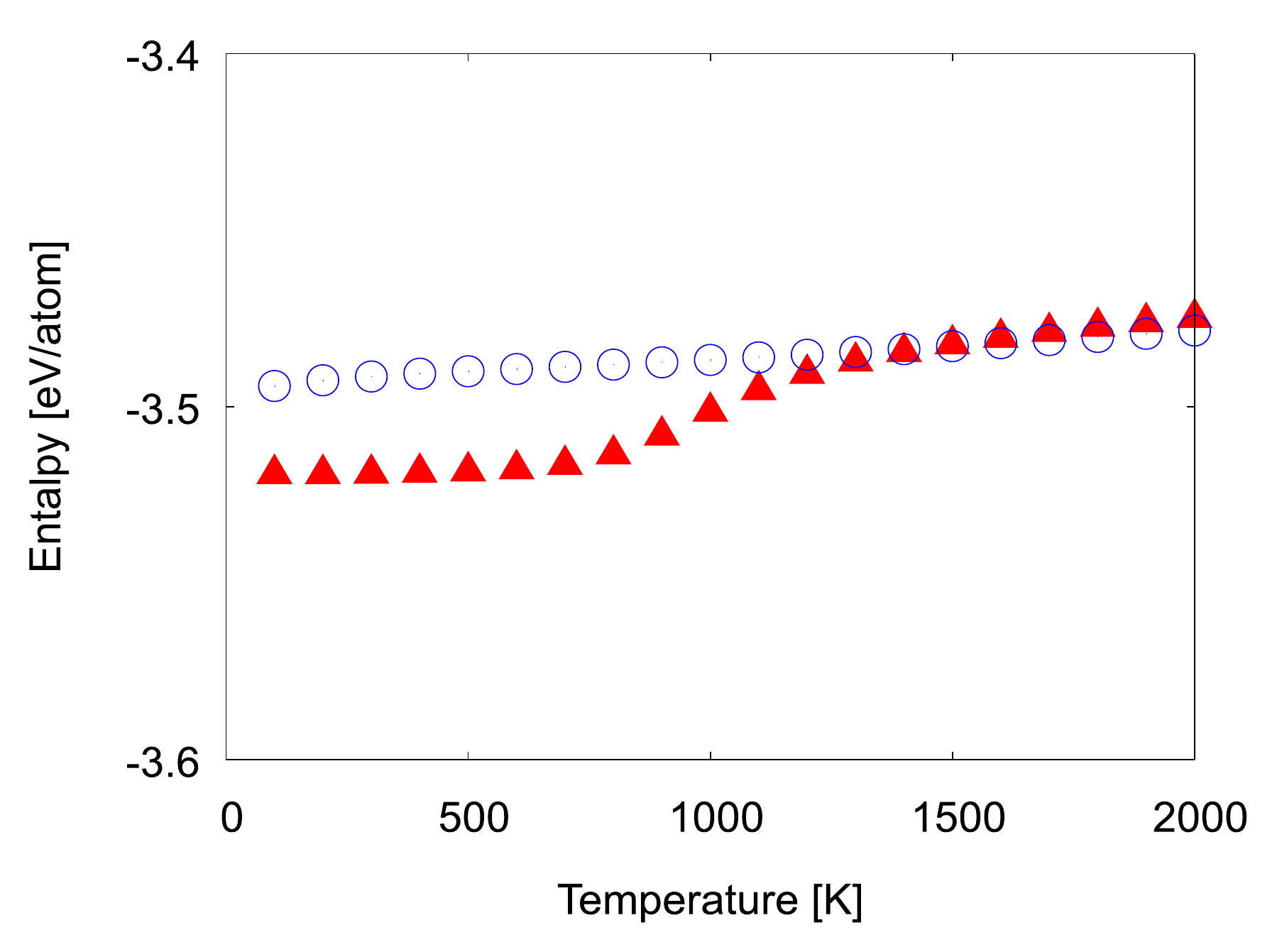}
\caption{Enthalpy for $\rm{Cu}_{16}\rm{Au}_{16}$. Closed triangles and open circles denote exact result from Eq.~(\ref{eq:exact}) and the result through present method.}
  \label{fig:internal_ene}%
\end{figure}

Below $T_{\rm c}$, the deviation is large because low energy states near ground state mainly contribute to configurational free energy but the product of the normal distribution function does not appropriately describe low energy states, i.e., in low $T$ we cannot neglect off-diagonal correlation coefficient matrix.
Here we show another consideration to explain the deviation below $T_{\rm c}$ in Fig.~\ref{fig:result}(a).
In Fig.~\ref{fig:internal_ene}, the enthalpy using our method is compared with that using exact calculation(Eq.~(\ref{eq:exact})). We confirm that the deviation in $F$ below $T_{\rm c}$ in Fig.~\ref{fig:result}(a) mainly comes from the deviation in enthalpy in Fig.~\ref{fig:internal_ene}.

The deviation at temperatures larger than $T_{\rm c}$ is around 0.7-0.9[meV/atom]. This accurate result has the probability to be accidentally good because system size is not large enough and boundary condition affects $\langle \xi_k \rangle$ which contributes to the center of configurational free energy. In Fig.~\ref{fig:R}, off-diagonal elements cannot be neglected when $N=32$. However Fig.~\ref{fig:R} shows that when system size increases off-diagonal elements and boundary condition can be neglected, thus error decreases in large system. This is also predicted by comparing the result between $N=32$ (Fig.~\ref{fig:result}(a)) and $N=16$ (Fig.~\ref{fig:result}(b)).

\section{\label{sec:level4} summary}
We propose a new approach to obtain configurational free energy directly and confirm that the present method is valid for disorder states at high temperature through comparison with first-principles-based thermodynamic simulation. Since the present method is based on the statistical information of lattice, we can evaluate configurational free energy using the expression of $Z$ that is separated into the the variables that depend on constituent elements and are independent of those.

\begin{acknowledgments}
This study was supported by a Grant-in-Aid for Young Scientists B (25820323) by JSPS and Research Grant from Hitachi Metals $\cdot$ Materials Foundation.
\end{acknowledgments}

\appendix

\section{\label{sec:appendixB} Structure integration when $x \neq 0.5$.}
Our approach is based on Eq.~(\ref{eq:decompose}), and this equation requires that \mbox{\boldmath $R$} is diagonal. However when $x \neq 0.5$ or in multicomponent system, \mbox{\boldmath $R$} is not always diagonal. In this situation, we should transform non-diagonal \mbox{\boldmath $R$} for $\xi$s to diagonal \mbox{\boldmath $\widetilde{R}$} for $\widetilde{\xi}$s (tilde denotes new expression through diagonalization.) in order to decompose the multivariate normal distribution function. Through this transformation, we get Eq.~(\ref{eq:SI}) in terms of $\widetilde{\xi}$s and $\widetilde{V_k}$s. In our approach, it is not essential what correlation functions express the atomic DOS, but essential that the atomic DOS is expressed by multivariate normal distribution function.

\begin{figure}
  \includegraphics[clip, width=\columnwidth]{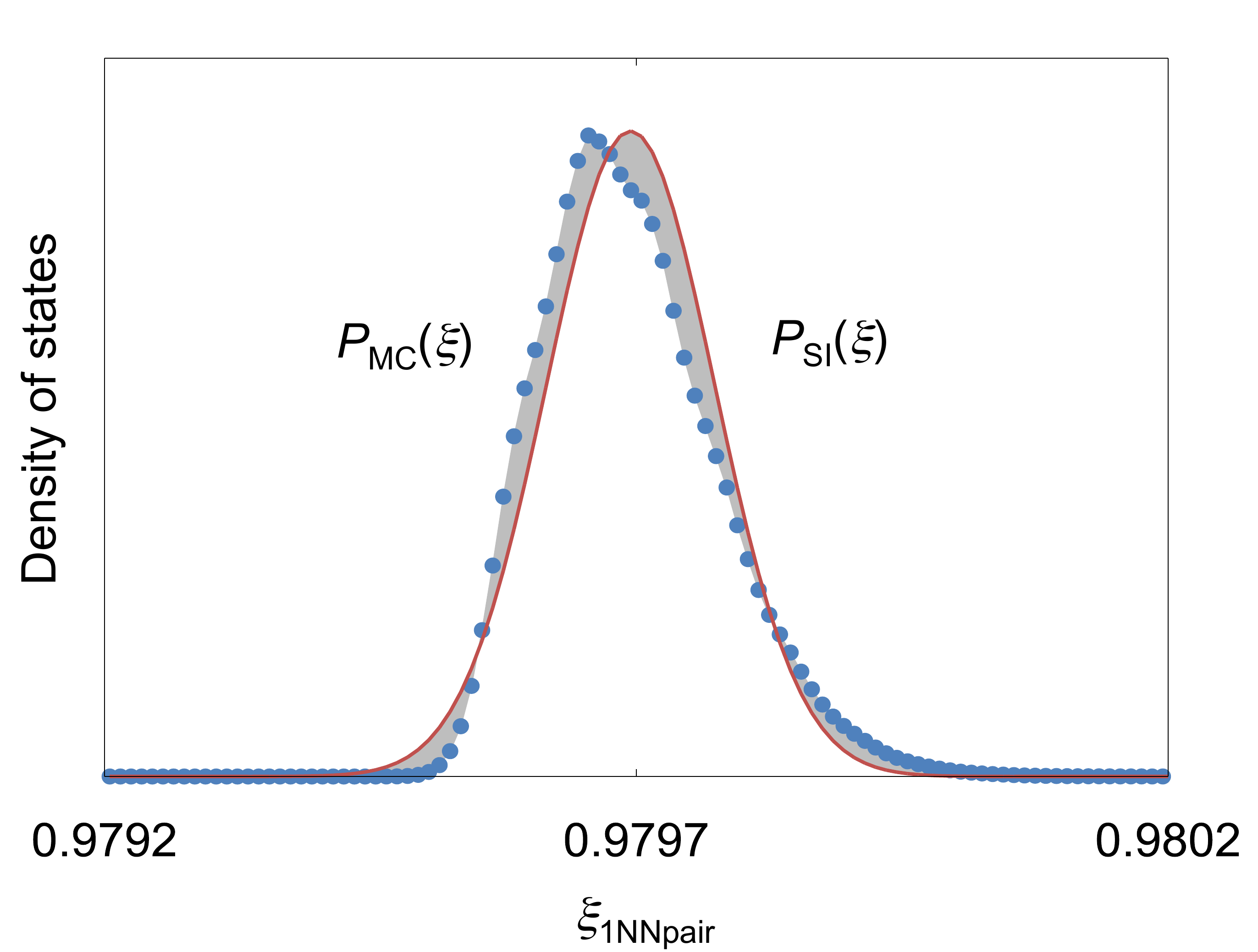}
  \caption{Points and solid curves denote the results of MC simulation in ${\rm A}_{56}{\rm B}_{10920}$ binary system on fcc lattice, $P_{\rm MC}$, and normal distribution function, $P_{\rm SI}$. Closed area shows the deviation of $P_{\rm MC}$ from $P_{\rm SI}$.}
  \label{fig:corr_dist_0056}%
\end{figure}

\begin{figure}
  \includegraphics[clip, width=\columnwidth]{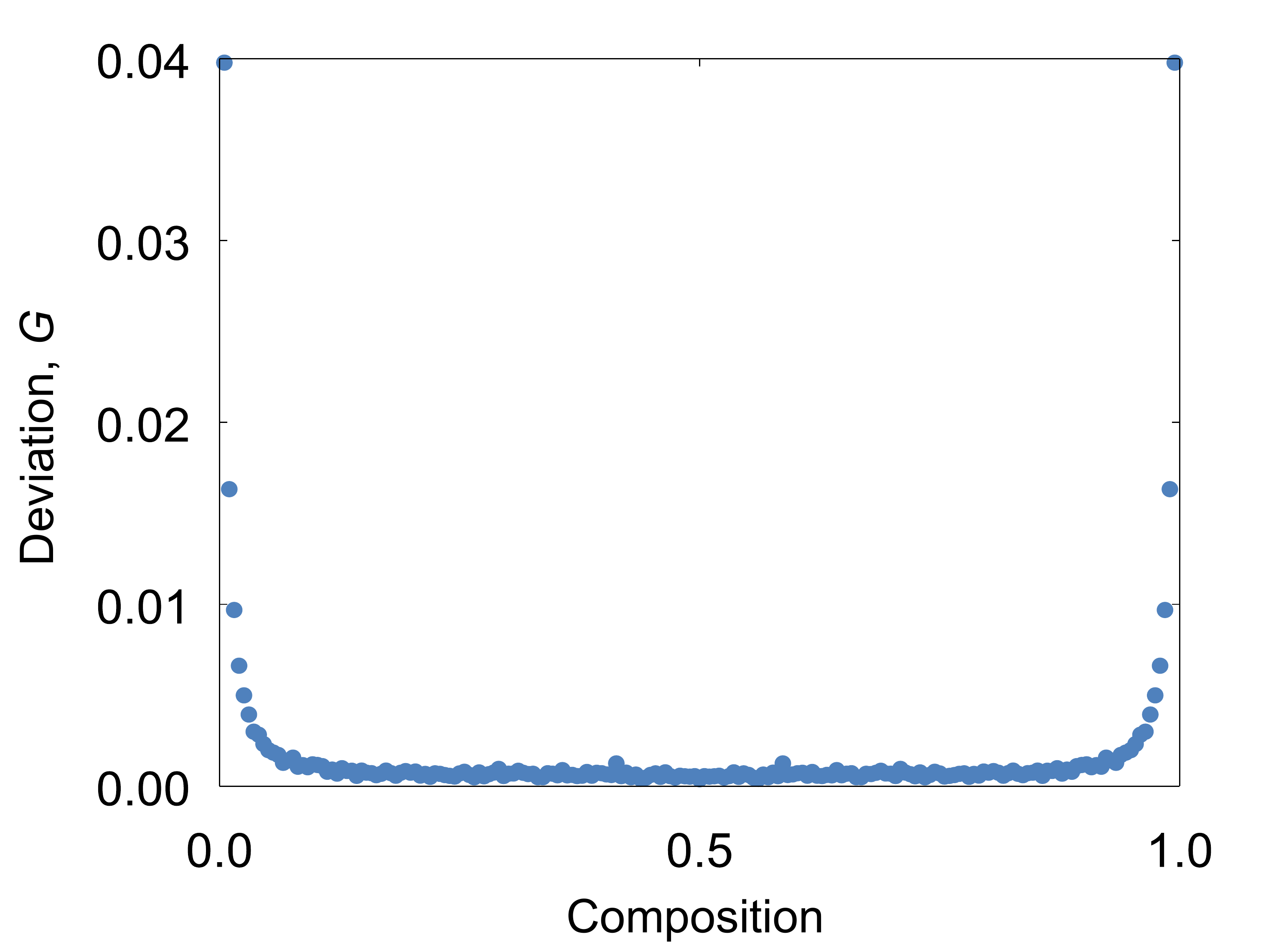}
  \caption{The deviation between $P_{\rm MC}$ and $P_{\rm SI}$, $G = \int |P_{\rm MC}(\xi) - P_{\rm SI}(\xi)| d\xi $, when $N=10976$. In Fig.~\ref{fig:corr_dist_0056}, The deviation is also shown as the example when $x \simeq 0.005$.}
  \label{fig:error_area}%
\end{figure}

When $x$ is far from 0.5, the DOS in terms of $\xi_k$ cannot be well expressed by the normal distribution function. This leads to significant error, therefore we should restricts the range of $x$ where our method is practically comparable to exact solution. In Fig.~\ref{fig:corr_dist_0056}, we show the example of how the DOS in terms of $\xi_k$ differs form normal distribution function through the MC simulation at $x \simeq 0.005$. The DOS through MC simulation, $P_{\rm MC}$, becomes like Poisson distribution, not normal distribution function, $P_{\rm SI}$. This deviation decrease with increasing $N$, therefore the suitable range of $x$ can change with the system size. Thus, in Fig.~\ref{fig:error_area}, under the condition of $N$ kept fixed, we show the deviations between $P_{\rm MC}$ and $P_{\rm SI}$ as the function of $x$ (e.g., defined as closed area in Fig.~\ref{fig:corr_dist_0056}). We find that the applicable $x$ range is between around 0.1-0.9 through practical simulation in $N \sim 10^4$.

\section{\label{sec:appendix} Single variate structure integration }
 In thermodynamic limit, the DOS in terms of correlation functions, $P(\xi)$, exhibits very sharp peak since the standard deviation of $P(\xi)$, $\sigma_\xi$, is proportional to $N^{-1/2}$. When $x=0.5$ and $\mu = 0$, it appears that the DOS describes the complete disorder state and corresponds to existing theories such as "ideal solution model" or "Bragg-Williams approximation". However, this is not true.

Here we first show the equivalence between $Z$ in terms of $E$ and $Z$ in terms of $\xi$ under the simple condition (but without lack of generality) when $E=V\xi$, i.e., $E$ is expressed by single correlation function. Since $V$ is proportional to $N$, we can rewrite
\begin{equation}
E = vN\xi ,
\end{equation}
where $v$ is $V$ per atom. Since $P(\xi)$ is normal distribution function of which $\sigma_\xi$ is equal to $1/\sqrt{ND}$, the DOS of $E$, $P(E)$, is also normal distribution function. Then, the variance of $P(E)$, $\sigma_E^2$, is
\begin{eqnarray*}
  \sigma_{E}^2  &=&   \langle E^2 \rangle - \langle E \rangle ^2 \\
  &=& v^2 N^2 \langle \xi^2 \rangle - v^2 N^2 \langle \xi \rangle ^2 \\
  &=& v^2 N^2 \sigma_\xi^2 ,
\end{eqnarray*}
where $\langle \rangle$ denotes the average of all possible states. It is easily shown that $Z= A \int P(E) \exp(-E/k_{\rm B}T)dE = A \int P(\xi)\exp(-vN\xi/k_{\rm B}T)d\xi$ through this transformation,
\begin{eqnarray*}
  Z &=& A \int P(E) \exp \left(-\frac{E}{k_{\rm B}T} \right)dE \\
  &=& A \int \frac{1}{\sqrt{2\pi \sigma_E^2}} \exp \left(-\frac{(E - \langle E \rangle)^2}{2 \sigma_E^2} - \frac{E}{k_{\rm B}T} \right) dE \\
  &=& A \int \frac{1}{\sqrt{2\pi v^2 N^2 \sigma_\xi^2}} \exp \left(-\frac{v^2 N^2(\xi - \langle \xi \rangle)^2}{2 v^2 N^2 \sigma_\xi^2} - \frac{vN\xi}{k_{\rm B}T} \right) vNd\xi \\
  &=& A \int \frac{1}{\sqrt{2\pi \sigma_\xi^2}} \exp \left(-\frac{(\xi - \langle \xi \rangle)^2}{2 \sigma_\xi^2} - \frac{vN\xi}{k_{\rm B}T} \right) d\xi \\
  &=&  A \int P(\xi) \exp \left(-\frac{vN\xi}{k_{\rm B}T} \right)d\xi .
\end{eqnarray*}

We also show that the temperature dependence of internal energy derived from structure integration in order to emphasize that our approach does not describe only the complete disorder state. The ensemble average, $\langle \rangle_Z$, is defined as:
\begin{equation}
  \langle E \rangle _ Z = \frac{A}{Z} \int E P(E) \exp \left(-\frac{E}{k_{\rm B}T} \right)dE .
\end{equation}
When we consider for instance $E=V\xi$, $E \ll k_{\rm B}T$ and $x = 1/2$, $\langle E \rangle _ Z$ is easily expressed by Taylor expansion,
\begin{eqnarray*}
   \langle E \rangle _ Z &\simeq& \frac{A}{Z} \int vN\xi P(\xi) \left(1-\frac{vN\xi}{k_{\rm B}T} \right)d\xi \\
   &=& \frac{A}{Z} \left( vN\langle \xi \rangle - \frac{v^2N^2\langle \xi^2 \rangle}{k_{\rm B}T} \right) \\
   &=& \langle E \rangle- \frac{v^2N^2\langle \xi^2 \rangle}{k_{\rm B}T} = \langle E \rangle - \frac{v^2N}{k_{\rm B}TD} .
\end{eqnarray*}
It is shown that at high $T$, $ \langle E \rangle _ Z \propto 1/T$, and this $T$ dependence certainly originates from disorder states modeled by $P(\xi)$, i.e., all possible microscopic states.

\bibliography{paper}

 \end{document}